%% file: typeN.tex
\documentclass[aps,prl,twocolumn,groupedaddress]{revtex4-2}

\usepackage{amsmath, slashed, amssymb}

\def\omicron{{o}}

\begin{document}

\begin{flushright}
\hspace{90mm} QMUL-PH-20-28 \\
MI-TH-2027
\end{flushright}


\title{The Weyl Double Copy for Gravitational Waves}

\author{Hadi Godazgar}
\email[]{hadi.godazgar@aei.mpg.de}
\affiliation{Max-Planck-Institut f\"ur Gravitationsphysik (Albert-Einstein-Institut), M\"uhlenberg 1, D-14476 Potsdam, Germany.}
\author{Mahdi Godazgar}
\email[]{m.godazgar@qmul.ac.uk}
\affiliation{School of Mathematical Sciences, Queen Mary University of London, Mile End Road, London E1 4NS, UK.}
\author{Ricardo Monteiro}
\email[]{ricardo.monteiro@qmul.ac.uk}
\affiliation{Centre for Research in String Theory, School of Physics and Astronomy,
Queen Mary University of London, E1 4NS, United Kingdom.}
\author{David Peinador Veiga}
\email[]{d.peinadorveiga@qmul.ac.uk}
\affiliation{Centre for Research in String Theory, School of Physics and Astronomy,
Queen Mary University of London, E1 4NS, United Kingdom.}
\author{C. N.\ Pope}
\email[]{pope@physics.tamu.edu}
\affiliation{George P. \& Cynthia Woods Mitchell Institute for Fundamental Physics and Astronomy, Texas A\&M University, College Station, TX 77843, USA.}
\altaffiliation{Also at Centre for Mathematical Sciences, Cambridge University, Wilberforce Road, Cambridge, CB3 0WA, UK.}

\date{October 7, 2020}

\begin{abstract}
We establish the status of the Weyl double copy relation for radiative solutions of the vacuum Einstein equations.  We show that all type N vacuum solutions, which describe the radiation region of isolated gravitational systems with appropriate fall-off for the matter fields, admit a degenerate Maxwell field that squares to give the Weyl tensor.  This relation defines a scalar that satisfies the wave equation on the background.  We show that for non-twisting radiative solutions, the Maxwell field and the scalar also satisfy the Maxwell equation and the wave equation on Minkowski spacetime. Hence, non-twisting solutions have a straightforward double copy interpretation.
\end{abstract}


\maketitle

The discovery of gravitational waves \cite{LIGO} one hundred years after Einstein formulated his general theory of relativity has led to an exciting new area of gravitational physics with possible important prospects for observational astrophysics; a development that has been anticipated eagerly for half a century \cite{gravastro}.  An important theoretical breakthrough in this direction will include an efficient and cost-effective method of generating gravitational wave templates; waveforms computed from the theory to be compared with observed waveforms \cite{tem0,tem01}; see \cite{num1,num2,tem1,tem2,tem3, tem4} for recent reviews.  Amongst the myriad approaches proposed to facilitate the easier and less time-consuming generation of templates is one \cite{Bern:2019nnu, Bern:2019crd} based on techniques adapted from string theory and supergravity scattering amplitude calculations, in particular the double copy method \cite{Kawai:1985xq, DC, DC1,Bern:2019prr}, which describes gravitational amplitudes as a kind of inner product of gauge theory amplitudes (hence ``double copy'').  

While initially found at the level of scattering amplitude relations, the double copy also exists at the level of classical solutions, including beyond perturbation theory for certain classes of spacetimes.  One class of solutions for which a double copy relation exists is (multi) Kerr-Schild solutions, which can be thought of as exact perturbative (around Minkowski) gravitational solutions \cite{Monteiro:2014cda, Luna:2015paa}. 
The correspondence between the double copy relations for scattering amplitudes and for classical solutions has been verified in various works \cite{Luna:2016due,Goldberger:2016iau,Luna:2016hge,Arkani-Hamed:2019ymq,Huang:2019cja,Kim:2019jwm,Luna:2020adi,Cristofoli:2020hnk}; see \cite{Saotome:2012vy,Neill:2013wsa} for earlier ideas in this direction.
Of particular interest in the present paper is the Weyl double copy relation that exists for vacuum type D solutions and pp-waves \cite{Walker, DR, lumonioc}. This relation is best expressed in spinor language \footnote{We summarise some important equations regarding spinor calculus in curved spacetime below.  For comprehensive introductions, see \cite{PenRind, stewart}.}. In the type D case, it can be shown that the Weyl spinor $\Psi_{ABCD} = (-2 \Phi^2)^{-1/4} \Phi_{(AB} \Phi_{CD)}$ with $\Phi_{AB}$ a non-degenerate Maxwell spinor and ${\Phi^2 \equiv \Phi^{AB} \Phi_{AB}}$.  Of particular significance is the fact that the Maxwell spinor also solves the Maxwell equation on Minkowski spacetime \cite{lumonioc}.  Furthermore, $\Phi^{1/2}$ solves the wave equation on Minkowski spacetime.  What lies behind these relations is the existence of the well-known hidden symmetry for type D vacuum solutions as expressed by the existence of a Killing 2-spinor \cite{carter,Walker}.  See \cite{Carrillo-Gonzalez:2017iyj, Ilderton:2018lsf,Lee:2018gxc,Berman:2018hwd,Andrzejewski:2019hub,Sabharwal:2019ngs,
Godazgar:2019ikr,Bah:2019sda,Alawadhi:2019urr,Banerjee:2019saj,Keeler:2020rcv,Bahjat-Abbas:2020cyb,Elor:2020nqe,Alawadhi:2020jrv,Alfonsi:2020lub,Adamo:2020qru} for related works.

In this paper, we extend the curved Weyl double copy relation to all type N vacuum 
solutions, which describe the radiation region of isolated gravitational 
systems. In particular, we show that $\Psi_{ABCD}=S^{-1} \Phi_{(AB} \Phi_{CD)}$ 
with $\Phi_{AB}$ a degenerate Maxwell spinor and $S$ some scalar 
that in particular satisfies the wave equation on the curved background.  For non-twisting radiative spacetimes, the Maxwell field and the scalar field also solve the Maxwell equation and the wave equation, respectively, on Minkowski spacetime.  This establishes the Weyl double copy in the sense of \cite{lumonioc} for this large class of spacetimes. Notice that, while the double copy for scattering amplitudes involves two copies of non-Abelian gauge theory, the first step in that procedure is to consider the double copy of the asymptotic states, which for linearised gauge theory are solutions to the Maxwell equation. The fact that certain exact gravity solutions can be interpreted as a double copy of a Maxwell field means that they should be interpreted as coherent states, an exact extension of the linearised asymptotic states in scattering amplitudes.
For twisting spacetimes, the Maxwell field and the scalar depend generically on the metric functions. Hence, they are solutions only on the curved spacetime. However, the standard double copy interpretation applies at the linearised level. This may be indicative of the fact that twisting solutions have an intrinsic non-Abelian nature. 


As a necessary step in extending the exact classical double copy tools to gravitational wave physics, we provide a systematic understanding of the status of the double copy for radiative solutions, beyond the most special example of pp-waves.  This study reveals interesting differences with the Weyl double copy for type D solutions.  In particular, the construction does not lead to a unique Maxwell field, since there is functional freedom associated to the scalar $S$.  
 In the cases where the Maxwell field can be thought of as living in Minkowski spacetime, i.e. for non-twisting solutions, it would be interesting to use novel approaches, e.g. \cite{Adamo:2017nia,Anastasiou:2018rdx,Kosower:2018adc}, to relate this new construction to the double copy for scattering amplitudes, as has been done for certain type D solutions.

\section{Spinor calculus}

The homomorphism between the Lorentz group and SL$(2,\mathbb{C})$ can be used to convert spacetime indices $\mu, \nu, \ldots$ into spinor $A, B, \ldots= \{1,2\}$ and conjugate spinor $\dot{A}, \dot{B}, \ldots= \{1,2\}$ indices, using the Van der Waerden matrices $\sigma^{\mu}_{A \dot{A}}$, which are constructed from the identity and Pauli matrices.  It is convenient to work in a spinor basis $\{\omicron^A, \iota^A\}$ with $\epsilon_{AB} o^A \iota^B = 1$.  In this basis, $\epsilon_{AB} = 2 \omicron_{[A} \iota_{B]}$ and can be used to lower indices $\psi_A=\psi^B\, \epsilon_{BA}\,. $
Similarly, $\psi^A=\epsilon^{AB}\, \psi_B.$  Associated with the spin basis is a null frame $(\ell,n,m,\bar{m})$ \footnote{The null vectors are constructed from the spinors as follows: $\ell^\mu \sim o_A \bar{o}_{\dot{A}}$, $n^\mu \sim \iota_A \bar{\iota}_{\dot{A}}$, $m^\mu \sim o_A \bar{\iota}_{\dot{A}}$ and $\bar{m}^\mu \sim \iota_A \bar{o}_{\dot{A}}$.}, so that
\begin{equation} \label{met:frame}
g_{\mu \nu} = 2 \ell_{(\mu} n_{\nu)} + 2 m_{(\mu} \bar{m}_{\nu)}.
\end{equation}
Our notation follows \cite{stephani}.
For a vacuum spacetime, the curvature is given by the Weyl tensor. Its spinorial version is fully determined by the totally symmetric Weyl spinor $\Psi_{ABCD}$ (and its complex conjugate), which satisfies the Bianchi identity
\begin{equation} \label{B}
 \nabla^{A \dot{A}} \Psi_{ABCD} = 0.
\end{equation}
Similarly, a solution of the Maxwell equation can be written in terms of a symmetric 2-spinor $\Phi_{AB}$ that solves
\begin{equation} \label{Max}
 \nabla^{A \dot{A}} \Phi_{AB} = 0.
\end{equation}
For type N solutions, choosing a spinor basis adapted to the principal null direction (PND) $\ell^\mu \sim o_A \bar{o}_{\dot{A}}$, the 
Newman-Penrose (NP) Weyl scalars, which correspond to various components of the Weyl spinor in the spinor basis, all vanish except ${\Psi_4=\Psi_{ABCD} \iota^A \iota^B\iota^C\iota^D = n^\mu \bar{m}^\nu n^\rho \bar{m}^\sigma C_{\mu \nu \rho \sigma}}$
and the Weyl spinor takes the simple form 
\begin{equation} \label{Weyl:N}
 \Psi_{ABCD} = \Psi_4\, \omicron_A \omicron_B \omicron_C \omicron_D.
\end{equation}

\section{Weyl double copy}

In spinor language, the curved background Weyl double copy relation is 
\begin{equation} \label{DC}
 \Psi_{ABCD}=\frac1{S}\, \Phi_{(AB} \Phi_{CD)},
\end{equation}
for some scalar $S$ and Maxwell spinor $\Phi_{AB}$.  Note that $\Phi_{AB}$ 
satisfies the Maxwell equation (\ref{Max}) in the fixed curved background 
metric, but it is viewed as a test field that does not back-react on the
geometry. From \eqref{Weyl:N}, it follows that the NP Maxwell scalars all
vanish except $\Phi_2$, and we have $\Phi_{AB} = \Phi_2\, o_{(A} o_{B)}$.
Thus the type N double copy relation is
\begin{equation} \label{DCN}
 \Psi_4=\frac1{S}\, (\Phi_2)^2 .
\end{equation}
The Maxwell 2-spinor is degenerate, which means that the electromagnetic field is null, i.e.\ the electric and magnetic fields are perpendicular and of equal magnitude.  An example of a null electromagnetic field is that of a plane electromagnetic wave in flat spacetime.  Now we must consider whether such a relation \eqref{DC} exists.  Expanding out the Bianchi identity \eqref{B} by substituting \eqref{Weyl:N} gives two equations:
\begin{gather} \label{bianchi:N}
 \omicron_A \nabla^{A \dot{A}} \log\Psi_4 + 
 4\, \omicron_A \iota^B \nabla^{A \dot{A}} \omicron_{B} - 
\iota_A \omicron^B \nabla^{A \dot{A}} \omicron_{B}=0
\end{gather}
and $\omicron_A \omicron^B \nabla^{A \dot{A}} \omicron_{B}=0.$  The second equation is equivalent to the statement that the null congruence generated by the PND is geodesic, $\kappa=0$, and shear-free, $\sigma=0$ \footnote{$\kappa = - \ell^\mu m^\nu \nabla_{\mu} \ell_\nu, \ \sigma = - m^\mu m^\nu \nabla_{\mu} \ell_\nu$}, which follow from the Goldberg-Sachs theorem \cite{goldberg}.  Expanding out the Maxwell equation in a similar fashion gives
\begin{equation} \label{max:N}
\omicron_A \nabla^{A \dot{A}} \log\Phi_2 + 
   2 \omicron_A \iota^B \nabla^{A \dot{A}} \omicron_{B} -  
\iota_A \omicron^B \nabla^{A\dot{A}} \omicron_B = 0,
\end{equation}
as well as the same equation above that is equivalent to $\kappa=\sigma=0.$  
Now, substituting $\Psi_4=(\Phi_2)^2/S$ 
into \eqref{bianchi:N} and simplifying this using \eqref{max:N} gives
\begin{equation} \label{S:N}
\omicron_A \nabla^{A \dot{A}} \log S - 
\iota_A \omicron^B \nabla^{A \dot{A}} \omicron_{B}=0.
\end{equation}
There is a clear structure in equations \eqref{bianchi:N}--\eqref{S:N}, where the coefficient of the middle term is the rank of the respective spinor. Equation \eqref{S:N}  translates, using $\bar{o}_{\dot{A}}$ and $\bar{\iota}_{\dot{A}}$, to 
\begin{equation} \label{Max:consis}
 \ell \cdot \nabla \log S - \rho = 0, \qquad m \cdot \nabla \log S - \tau = 0,
\end{equation}
where $\rho$ and $\tau$ are NP spin coefficients \footnote{$\rho = - \bar{m}^\mu m^\nu \nabla_{\mu} \ell_\nu, \ \tau = - n^\mu m^\nu \nabla_{\mu} \ell_\nu$}.  $\rho$ parametrises the expansion and twist of the null congruence generated by $\ell$, while $\tau$ parametrises the transport of $\ell$ along the flow generated by $n$.  A simple 
calculation shows that the integrability condition on the equations \eqref{Max:consis} is satisfied, which means that 
they are simple integral equations that can always be solved.  
Thus, we are guaranteed the existence of a scalar $S$ satisfying these 
equations, which then gives a Maxwell field 
$\Phi_2=\sqrt{\Psi_4\, S}$.  In tensor language, this Maxwell 
spinor translates to a field strength (called the `single copy') of the form 
\begin{equation}
F = \Phi_2\, \ell^{\flat} \wedge m^{\flat} + 
\bar{\Phi}_2\, \ell^{\flat} \wedge \bar{m}^{\flat},\label{FPhi}
\end{equation}  
where $\ell^\flat$ denotes the 1-form $\ell^\flat = \ell_\mu dx^\mu$, and similarly for ${m}^{\flat}$ and $\bar{m}^{\flat}$.
This establishes the curved Weyl double copy for type N vacuum solutions.  

Furthermore, it is simple to show using \eqref{S:N} that $S$ solves the wave equation
\begin{equation}
\Box S = \nabla_{A\dot{A}} \nabla^{A\dot{A}} S = 2 \omicron_A \iota_B \nabla^A{}_{\dot{A}}\nabla^{B \dot{A}}S= 0.
\end{equation}
The real scalar field in the double copy construction (called the `zero-th copy') is the real part of $S$.

These results mirror those that exist for type D solutions.  In order to investigate whether the Maxwell field and the scalar field also satisfy the equations of motion on Minkowski spacetime, we investigate the different classes of type N solutions in turn.

\section{Type N vacuum solutions}
Type N vacuum solutions are classified in terms of the optical properties of the congruence generated by the PND, i.e.~by the values of the optical scalars; see e.g. \cite{stephani}. We have $\kappa=\sigma=0,$ as mentioned before; the properties that remain are parametrised by the spin coefficient
$\rho = -(\Theta + i\, \omega)$, where $\Theta$ denotes the expansion of the congruence and $\omega$ denotes its twist.  The different cases lead to three distinct classes of solutions:
\begin{itemize}
 \item Kundt solutions: $\Theta = 0,$ which implies that ${\omega=0}$ \footnote{Substituting $\Theta=\sigma=R_{\mu \nu}=0$ into the Raychaudhuri equation $\ell \cdot \nabla \Theta - \omega^2 + \Theta^2 + \sigma \bar{\sigma} + \frac{1}{2} R_{\mu \nu} \ell^\mu \ell^\nu=0$ gives $\omega=0$.}.
 \item Robinson-Trautman solutions: $\Theta \neq 0, \ \omega=0.$
 \item Twisting solutions: $\Theta \neq 0, \ \omega \neq 0.$
\end{itemize}
Choosing a null frame for which $\ell$ is the PND, so that 
$\Psi_0=\Psi_1=\Psi_2=\Psi_3=0$, we consider each case separately.

\subsection{Kundt solutions}
There are two kinds of type N Kundt solutions, both corresponding to plane-fronted wave solutions \footnote{See Theorem 31.2 of \cite{stephani}.}.  Plane-fronted waves with parallel propagation (pp-waves) are given by the metric
\begin{equation} \label{pp}
 ds^2 = - 2 du\, (dv + H du) + 2 dz d\bar{z},
\end{equation}
with $H(u,z,\bar{z}) = f(u,z) + \bar{f}(u,\bar{z})$ for general 
functions $f$. Choosing
\begin{equation}
\label{tetradppwave}
\ell=\partial_v,\quad 
n = \partial_u - H\, \partial_v,\quad m=\partial_z,
\end{equation}
one has 
$\rho=\tau=0$ and so (\ref{Max:consis}) implies 
$S= S(u,\bar{z}),$ while the Weyl scalar 
$\Psi_4= \partial_{\bar{z}}^2 \bar{f},$ so (\ref{DCN}) implies that 
\begin{equation}
\Phi_2= \sqrt{\partial_{\bar{z}}^2 \bar{f}\, S(u,\bar{z})}.
\end{equation}

The other class of plane-fronted waves is given by
\begin{equation}
 ds^2 = - 2 du \Big(dv +W dz + \bar{W} d\bar{z} + H du \Big) + 2 dz d\bar{z},
\end{equation}
with $ W(v,z,\bar{z})= - 2v \,(z+\bar{z})^{-1}$ and
$$H(u,v,z,\bar{z}) = \left[f(u,z) + \bar{f}(u,\bar{z})\right](z+\bar{z}) - \frac{v^2}{(z+\bar{z})^2};$$
again $f(u,z)$ is arbitrary.  Choosing 
$$ \ell=\partial_v, \ n = \partial_u - (H+ W \bar{W})\, \partial_v + \bar{W} \partial_z + W \partial_{\bar{z}}, \ m=\partial_z, $$
one has $\rho = 0$, $\tau = 2\beta= -(z+\bar{z})^{-1},$ so (\ref{Max:consis}) 
gives ${S = \zeta(u,\bar{z})/(z+\bar{z}).}$  The Weyl scalar 
$\Psi_4= (z+\bar{z})\, \partial_{\bar{z}}^2 \bar{f}$, so (\ref{DCN}) 
implies that 
\begin{equation}
\Phi_2= \sqrt{\partial_{\bar{z}}^2 \bar{f}\, \zeta(u,\bar{z})}.
\end{equation}

Given that the only non-zero components of $F_{\mu \nu}$ are for $\mu \nu = [uz]$ and $[u\bar{z}]$, the simple  form of the relevant components of $g^{\mu \nu}$ and the fact that $g=1$ give
\begin{align}
\nabla_{\nu} F^{\mu \nu} &= \frac{1}{\sqrt{|g|}} \partial_\nu \left( \sqrt{|g|}\,  g^{\mu \rho} g^{\nu \sigma} F_{\rho \sigma}  \right)  \notag \\
&=  \partial_\nu \left( \eta^{\mu \rho} \eta^{\nu \sigma} F_{\rho \sigma}  \right) = 0.
\label{Max:flat}
\end{align}
On the other hand, $S$ does not depend on $f(u,z)$ or $\bar{f}(u,\bar{z})$, meaning that it must solve the wave equation on any member of the family. 
In particular, it solves the wave equation on Minkowski spacetime.
This implies that the Maxwell and the scalar fields also satisfy their equations on Minkowski spacetime, establishing the Weyl double copy for type N Kundt solutions.

\subsection{Robinson-Trautman solutions}
Type N Robinson-Trautman solutions take the 
form \footnote{See Theorem 28.1 and Section 28.1.2 of \cite{stephani}.}
\begin{equation} \label{RT}
 ds^2 = - H\, du^2 - 2 du\, dr + \frac{2 r^2}{P^2}\, dz\,d\bar{z}\,,
\end{equation}
with $H(u,r,z,\bar{z}) = k - 2 r\, \partial_u\log P\ 
(\hbox{where}\ k = 0,\pm 1)$ and $2 P^2\partial_z\partial_{\bar z}  
\log P(u,z,\bar{z}) = k$.  Choosing
\begin{equation}
\ell = \partial_r, \quad n = \partial_u - {\textstyle{\frac{1}{2}}} H \partial_r, \quad m = -\frac{P}{r}\, \partial_z,
\end{equation}
one has $ \rho = -r^{-1}, \ \tau = 0,$ so (\ref{Max:consis}) gives 
$S = - \zeta(u,\bar{z})/r.$  Now ${\Psi_4 = 
- \frac{P^2}{r} \partial_u \left( \frac{\partial_{\bar{z}}^2 P}{P} \right)}$, 
so (\ref{DCN}) determines that
\begin{equation}
\Phi_2= \frac{P}{r}\, \sqrt{ \zeta(u,\bar{z})\, 
\partial_u \left( \partial_{\bar{z}}^2 P/P \right)}.\label{RTzeta}
\end{equation}

  As an example, consider Robinson-Trautman solutions with $k=0$ in (\ref{RT}).
Writing $P=e^W$ we have $\partial_z\partial_{\bar z}\, W=0$ and hence
$W=w(u,z) + \bar w(u,\bar z)$, implying that $\Psi_4=-P^2/r\, \partial_u[
\partial_{\bar z}^2 \bar w(u,\bar z) + (\partial_{\bar z} \bar w(u,\bar z)^2]$.
We can obtain type N solutions of the Maxwell equation in the Robinson-Trautman
background by taking
\begin{equation}
A = \gamma(u,z,\bar z)\, du,
\end{equation}
where $\partial_z\partial_{\bar z}\, \gamma=0$ and hence $\gamma=
h(u,z) +\bar h(u,\bar z)$. Thus from (\ref{FPhi}) we have $\Phi_2 = 
-P/r\, \partial_{\bar z}\, \bar{h}(u,\bar z)$.  Plugging into (\ref{DCN}) we have
\begin{equation}
\partial_u [\partial_{\bar z}^2 \bar w(u,\bar z) +
          \big(\partial_{\bar z}\bar w(u,\bar z)\big)^2] =-\frac{1}{rS}\,
     \big(\partial_{\bar z} \bar h(u,\bar z)\big)^2,
\end{equation}
and so indeed we have that $S= -\zeta/r$, where $\zeta$ is a function
only of $u$ and $\bar z$, as required in the general result stated above.

As with Kundt solutions, the only non-zero components of $F_{\mu \nu}$ are for 
$\mu \nu = [uz]$ and $[u\bar{z}]$.  As before, using the fact that $\sqrt{|g|} = r^2/P^2$ 
and the relevant components of $g^{\mu \nu}$, it can be shown that \eqref{Max:flat} holds. 
Once again, $S$ is independent of $P$ and solves the wave equation on any member of 
the family \eqref{RT}, including Minkowski. Hence, both $F_{\mu\nu}$ and $S$ 
satisfy their equations also on the flat background,
 establishing the Weyl double copy for Robinson-Trautman solutions.

\subsection{Twisting solutions}

Type N solutions with non-vanishing twist are more complicated, with only 
one explicit solution known \cite{hauser}.  The general metric is given 
by \footnote{See Chapter 29 of Ref.\ \cite{stephani}}
\begin{align}
 ds^2 = - 2 (du &+ L dz + \bar{L} d \bar{z}) \Big[dr + W dz + \bar{W} d \bar{z}  \\ \notag
 &+ H \left( du + L dz + \bar{L} d \bar{z} \right)\Big] + \frac{2}{P^2 |\rho|^2} dz d\bar{z},
\end{align}
\begin{align*}
  &\rho^{-1}=- (r + i \Sigma), \quad 2 i\, \Sigma(u,z,\bar{z}) = P^2 (\bar{\partial} L - \partial \bar{L}), \\
 &W(u,r,z,\bar{z}) = \rho^{-1}\, \partial_u L + i \partial \Sigma, \quad \partial = \partial_z - L \partial_u,  \notag \\
 &H(u,r,z,\bar{z}) = \frac{1}{2} K - r \partial_u \log P, 
\end{align*}
with $K = 2 P^2 \Re \left[ \partial(\bar{\partial} \log P - 
\partial_u \bar{L}) \right].$  There exists a residual gauge freedom to 
choose $P=1$, but we shall not yet impose this choice.  The solution 
is determined by the complex scalar $L$, which satisfies
\begin{equation*}
 \Sigma K + P^2 \Re \left[ \partial \bar{\partial} \Sigma - 2 \partial_u \bar{L} \partial \Sigma - \Sigma \partial_u \partial \bar{L} \right] = 0,
 \quad \partial I = 0,
\end{equation*}
and $\partial_u I \neq 0$, with $I = \bar{\partial} (\bar{\partial} \log P 
- \partial_u \bar{L}) + (\bar{\partial} \log P - \partial_u \bar{L})^2.$  
Choosing
\begin{equation}
\ell = \partial_r, \quad n = \partial_u - H\, \partial_r, \quad m = -P \bar{\rho} \, (\partial - W \partial_r),
\end{equation}
$\rho$ is as defined above, while $\tau=0.$  Equation (\ref{Max:consis}) then
implies that 
$S = \rho \, \chi(u,z,\bar{z})$, with $\chi$ satisfying
\begin{equation}
  \partial \chi - \partial_u L\, \chi = 0.\label{delzetaeqn}
\end{equation}
Defining new coordinates $(v,w)=(I,z)$, the above equation can be solved 
using the method of characteristics ($I = \textup{constant}$ correspond to 
the characteristics)
\begin{equation} \label{twist:zeta}
 \chi(v,w) =  \zeta(I)\, e^{\int  \left[ \left( \frac{\partial I(u,z)}{\partial u}\right)(v,w') \times \frac{\partial L(v,w')}{\partial v} \right] d w'},
\end{equation}
with $\zeta(I)$ arbitrary.  The Weyl scalar $\Psi_4 = 
\rho\, P^2\, \partial_u I$, and so (\ref{DCN}) implies
\begin{equation}
{\Phi_2= \rho\, P\, \sqrt{ \partial_u I\,\chi(u,z,\bar{z})} }.
\end{equation}

  Only one twisting type N solution, found by Hauser \cite{hauser},
is known explicitly.  The metric functions are given by
$$
P=(z+\bar z)^{3/2}\, f(t),\quad t\equiv \frac{u}{(z+\bar z)^2},\quad
L=2i (z+\bar z),
$$
where $f$ satisfies $16(1+t^2) f''(t) +
 3f(t)=0$, which is a hypergeometric equation, 
and $I$ turns out to be given by
\begin{equation}
I=\frac{3}{2[(z+\bar z)^2 - iu]}.
\end{equation}
The solution to \eqref{Max:consis} is
\begin{equation}
S=\rho\, \zeta(I)~, 
\end{equation}
where $\zeta(I)$ is arbitrary. As expected, this is consistent with the general result \eqref{twist:zeta}. 
The Weyl scalar is $\Psi_4= (2i/3)\, \rho P^2\, I^2$, implying that 
\begin{equation}
\Phi_2=\rho\,P\,I \sqrt{\frac{2\,i\,\zeta(I)}{3}}~. \label{eq: Phi2 hauser}
\end{equation}

As a further remark about the twisting type N solutions, we note that
if the gauge freedom to set $P=1$ is employed, the metric is specified purely
in terms of the function $L(u,z,\bar z)$, and the type N and Ricci flat
conditions may be succinctly condensed down to just
\begin{equation}
\partial I=0, \quad \Im(\bar\partial \bar\partial\partial\partial L)=0,
\quad \hbox{where}\quad I=-\partial_u\bar\partial\bar L.
\end{equation}
The Weyl curvature is given by $\Psi_4=\rho\,\partial_u I$.  

In contrast to non-twisting solutions, the second equality in \eqref{Max:flat} does not hold for twisting solutions. Therefore, while there is a curved Weyl double copy relation, in this case it does not translate to a relation where the Maxwell field and the scalar can be thought of as Minkowski fields, unless we consider all the fields (gravity, Maxwell and scalar) at the linearised level.

\section{Non-uniqueness}

In all the cases above, neither the Maxwell field nor the scalar field are uniquely determined. They are fixed only up to an arbitrary function of some of the coordinates, which we are free to choose. This contrasts with the Weyl double copy for vacuum type D solutions, for which, in a spinor basis adapted to the principal null directions, we have $S^3\propto(\Phi_2)^{3/2}\propto \Psi_4$, where the proportionality is up to complex parameters \cite{lumonioc}; hence the Maxwell and scalar fields are functionally fixed. This feature is related to the fact that vacuum type D spacetimes are fully determined up to a few parameters, whereas vacuum type N spacetimes (of any class, as seen above) have functional freedom. By analogy, there is additional freedom in the Maxwell and scalar fields in the curved background. 

In considering a special choice, we may ask whether it is possible to choose $\Phi_2$ and $S$ to be given by specific powers of $\Psi_4$, as in the type D case, i.e.\ there exists some constant $a$ such that $\Phi_2\propto(\Psi_4)^a$ and $S\propto (\Psi_4)^{2a-1}$.  The functional dependence of the results above implies that this possibility holds only for Kundt solutions.  For pp-waves, the power is actually undetermined, i.e.\ the relation above holds for any $a$.  A simple choice is ${a=1/2}$, where $S$ is constant, and in fact this choice implies that Maxwell plane waves double copy to gravitational plane waves ($\Phi_2$ and $\Psi_4$ are functions of $u$ only).  For the other plane-fronted Kundt solutions, such a relation is possible for $a=0$, in which case $S\propto (\Psi_4)^{-1}$.  Analogously simple choices for the other type N classes are: $S\propto 1/r$ for Robinson-Trautman solutions and $S\propto \rho$ for twisting solutions.

Interestingly, pp-waves are the only type N solutions admitting a Killing 2-spinor \cite{DR}, another feature that they share with type D solutions.

A twistorial version of the Weyl double copy is given in \cite{White:2020sfn}, focusing on type D but also introducing some type III cases, at least at the linearised level. It would be interesting to study whether this twistorial version explains the non-uniqueness of the type N Weyl double copy found here.
\\

\begin{acknowledgments}
\noindent
{\textbf{Acknowledgements}} We would like to thank Donal O'Connell and Chris White for discussions.  We thank Pujian Mao for pointing out an incorrect statement in the abstract of a previous version of the paper. HG would like to thank Queen Mary University of London for hospitality during the course of this work.  HG is supported by the ERC Advanced Grant Exceptional Quantum Gravity (Grant No.\ 740209).  MG and RM are supported by Royal Society University Research Fellowships.  CNP is partially supported by DOE grant DE-FG02-13ER42020. DPV is supported by a Royal Society studentship grant.

\end{acknowledgments}

\input{typeN.bbl}

\end{document}

%% file: typeN.bbl
%